# Giant photoinduced lattice distortion in oxygen-vacancy ordered thin films

*Bingbing Zhang[1*], Xu He[2], Jiali Zhao[1], Can Yu[1], Haidan Wen[3], Sheng Meng[4], Eric Bousquet[2], Yuelin Li[3], Kuijuan Jin[4], Ye Tao[1*] and Haizhong Guo[5*]*

[1]Institute of high energy physics, Chinese academy of sciences, Beijing, 100049, China

[2] Physique Théorique des Matériaux, Q-MAT, CESAM, Université de Liège, B-4000 Liège, Belgium

[3]Advanced Photon Source, Argonne National Laboratory, Argonne, Illinois, 60439, USA

[4]Institute of Physics, Chinese Academy of Sciences, Beijing, 100190, China

[5]School of Physical Engineering, Zhengzhou University, Zhengzhou, 450001, China

*zhangbb@ihep.ac.cn
*taoy@ihep.ac.cn
*hguo@zzu.ed

## Abstract

Despite of the tremendous efforts spent on the oxygen vacancy migration in determining the property optimization of oxygen-vacancy enrichment transition metal oxides, few has focused on their dynamic behaviors non-equilibrium states. In this work, we performed multi-timescale ultrafast X-ray diffraction measurements by using picosecond synchrotron X-ray pulses and femtosecond table-top X-ray pulses to monitor the structural dynamics in the oxygen-vacancy ordered $SrCoO_{2.5}$ thin films. A giant photoinduced strain ($\Delta c/c > 1\%$) was observed, whose distinct correlation with the pump photon energy indicates a non-thermal origin of the photoinduced strain. The sub-picosecond resolution X-ray diffraction reveals the formation and propagation of the coherent acoustic phonons inside the film. We also simulate the effect of photoexcited electron-hole pairs and the resulting lattice changes using the Density Function Theory method to obtain further insight on the microscopic mechanism of the measured photostriction effect. Comparable photostrictive responses and the strong dependence on excitation wavelength are predicted, revealing a bonding to anti-bonding charge transfer or high spin to intermediate spin crossover induced lattice expansion in the oxygen-vacancy films.

Multivalent transition-metal oxides (TMOs) have claimed serious attention on basics of their intriguing physical properties and potential applications in energy technologies [1,2]. Among TMOs, the oxygen deficient SrCoOx stands out as a prime candidate for studying the role of oxygen stoichiometry in determining its physical behaviors due to the existence of two reversible topotactic phases [3-6]. The long-range ordering oxygen vacancies and rich polyhedral configurations make $SrCoO_{2.5}$ (SCO) very attractive in both applications and fundamental researches [4, 7]. The ordered oxygen vacancies in SCO result in the super-tetragonality of the Co-O tetrahedron layer, which suppresses the tilting of Co-O octahedron layer, like the ferroelectrics like $BaTiO_3$, $PbTiO_3$, where the tilting of octahedron is usually suppressed due to the competition with polarization [8]. Various exotic properties have been reported in SCO by controlled modification of the ambient condition including strain [9,10], pressure [11], voltage [12,13], and heat treatment [4,6]. However, few reports have focused on the dynamic behaviors under nonequilibrium states. The ultrafast X-ray diffraction (UXRD) technique has exemplified its power in interrogating transient phenomena upon photoexcitation in different nanostructures of functionalities such as $Sr_2IrO_4$ [14], $PbTiO_3$ [15], and $BiFeO_3$ [16,17].

Here we present the UXRD experiments on SCO with both picosecond and femtosecond temporal resolution on the SCO thin films in combination with first-principles calculations. The ultrafast structural dynamics is monitored upon excitation by different photon energies (1.55 eV and 3.1 eV). The giant out-of-plane strain ($\Delta c/c >$ 1%) upon 3.1 eV photoexcitation is observed, which is significantly larger than that under 1.55 eV photoexcitation. The calculation results also indicate a large photostriction effect and strong dependence on excited photon energies, in good agreement with the experimental data. We attribute the effect to photoexcited charge transfer that couples strongly to the SCO unit cell and drives a transient lattice distortion.

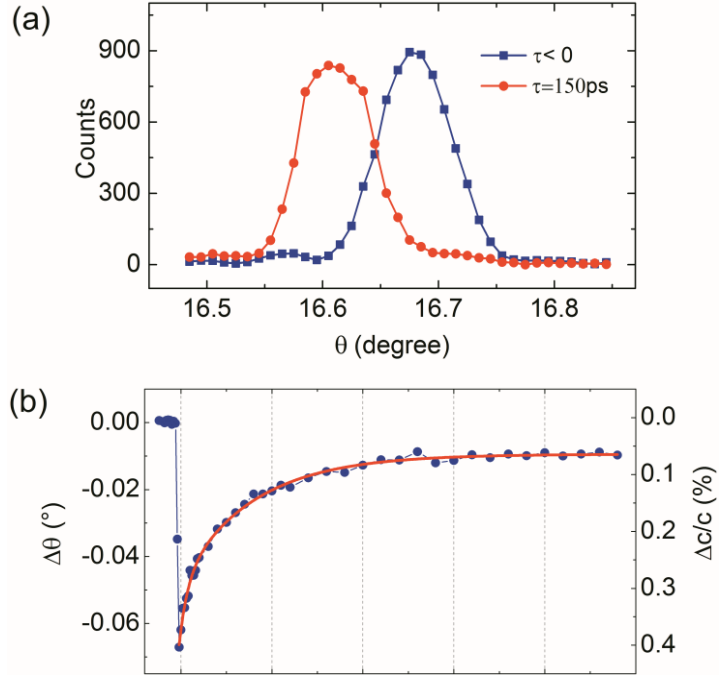

FIG. 1. (a) The θ-2θ scans of the SCO (008) reflection before (τ < 0) and after (τ = 150 ps) the excitation of 3.1 eV laser pulses with incident fluence of 1.55 mJ/cm$^2$. (b) The extracted peak shift and corresponding strain is empirically fitted to a biexponential decay function. No peak width change was evolution can be found.

We investigated the brownmillerite structure SCO, with alternating Co-O octahedral and tetrahedral layers stacked along the [001] direction and ordered oxygen vacancy channels arranged in the a-b plane, grown on (001) LaAlO$_3$ substrates. The direct band gap (~2.18 eV) and Mott gap (~ 0.45 eV) of the SCO film have been reported [3] and can be derived from our optical absorption data (Fig. S1).

The time-resolved X-ray diffraction experiments upon photoexcitation of 1.55 eV (800 nm) and 3.1 eV (400 nm, frequency doubling of 800 nm) were performed on a 43-nm-thick SCO film at 7-ID-C beamline of Advance Photon Source. The transient rocking curves were probed by X-ray pulses with photon energy of 11 keV after excitation of 3.1 eV laser pulses. The penetration depth of 3.1 eV laser in SCO is measured to be 30 nm from the optical absorption. The angular shift of the smaller Bragg angle shown in Fig. 1(a) corresponds to an ultrafast out-of-plane expansion Δc/c = 0.4 % for the incident fluence of 1.55 mJ/cm$^2$. Comparing with the previous BiFeO$_3$ work (Δc/c = 0.41 % [17] at absorbed fluence of 3.2 mJ/cm$^2$), much smaller fluence was needed to achieve a comparable strain (absorbed fluence: 0.87 verses 3.2 mJ/cm$^2$).

The evolution of angular shift (Fig. 1(b)) and peak width change (Fig. 1(c)) of the SCO (008) reflection as a function of time are extracted from the time-resolved rocking curves. The angular shift change can be well fitted by a biexponential decay function with an offset, which shows a fast time constant of 0.45 ns and a slow time constant of 4.3 ns. Unlike the giant change of the Bragg angle, the peak width of the reflection remains unchanged after photoexcitation, indicating homogeneous spatial strain profile after the optical excitation.

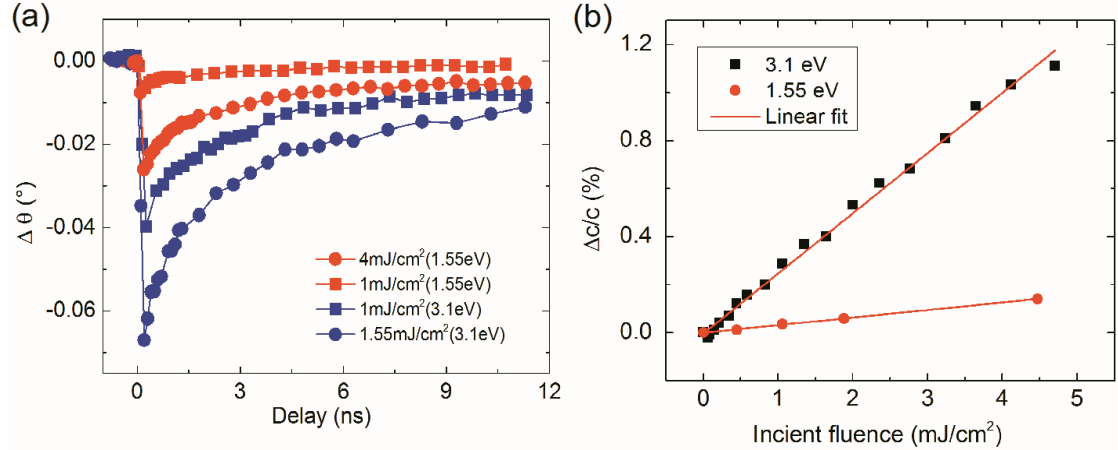

FIG. 2. (a) Angular shift of SCO (008) reflection as a function of time delay at the different pump fluence and photon energies. (b) Photoinduced strain at $\tau$ = 150 ps as a function of incident laser fluence from 0 to 5 mJ/cm$^2$ upon excitation of 3.1 eV and 1.55 eV laser pulses, together with a linear fit to the data.

The photoinduced strains at various pump photon energies and incident fluence were also measured, as shown in Fig. 2(a). The angular shifts of the (008) Bragg peak excited by 1.55 eV pump pulse are much smaller than the 3.1 eV case, as shown in Fig 2(b). Taking the two 1 mJ/cm$^2$ cases, for example, the maximum angular shift at $\tau$ = 150 ps for the 3.1 eV case is about 6 times larger than that in 1.55 eV case. However, the actual absorption fluence for 3.1 eV and 1.55 eV are 0.675 mJ/cm$^2$ and 0.41 mJ/cm$^2$, respectively. The difference is not enough to explain the strong photon energy dependence but reminiscent of the two featured peaks in the optical absorption data in Fig S1. Hence, we might ascribe this large difference in photostriction to two distinct photoexcited charge transfers, the intraband *d-d* transition ($\alpha$ peak) and the interband *p-d* transition ($\beta$ peak) [3]. Explicit linear fluence dependence of angular shift at $\tau$ = 150 ps was observed for both pump photon energies (Fig. 2(b)), indicating strains of up to 1%. For fluence bigger than 5 mJ/cm$^2$, the strain reaches saturation together with apparent sample degradation.

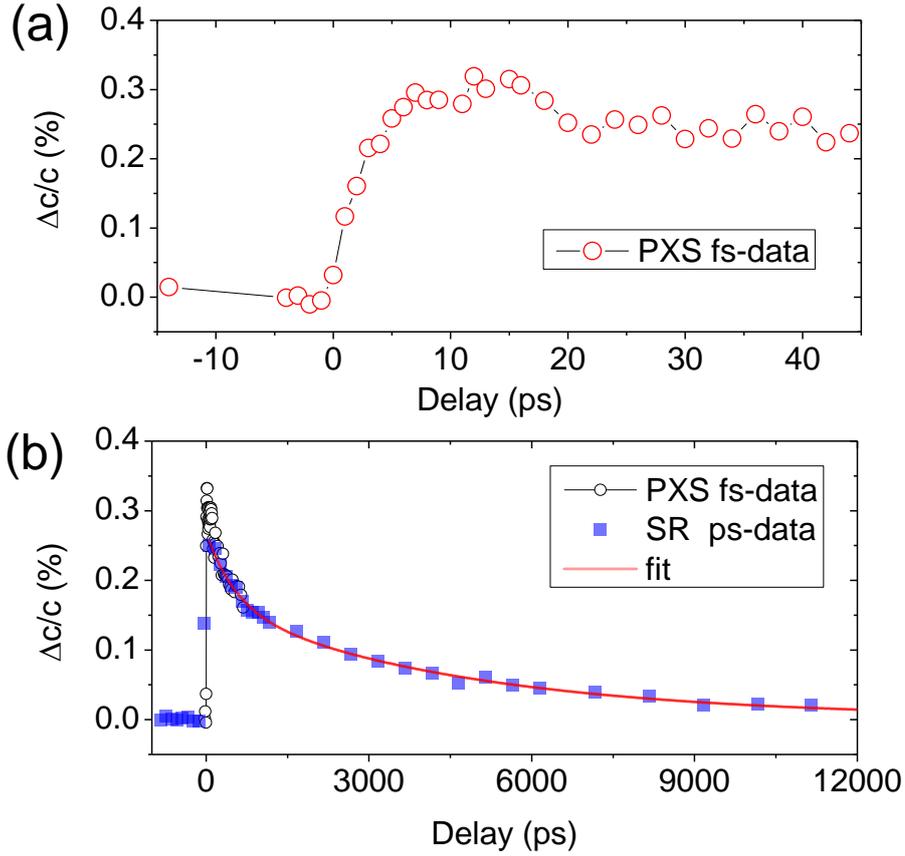

FIG. 3. (a) Transient angular shift of the (008) reflection of the SCO film, as well as the corresponding photoinduced strain versus time delay from -10 ps to 45 ps. (b) The photoinduced strain by merging the fs and ps UXRD data together.

To provide insight into the ultrafast buildup of the strain right after excitation, we performed sub-ps UXRD measurements on the same sample using a laser-based Plasma X-ray Source. The photoinduced strain as a function of delay within first 600 ps upon 400 nm laser pump at fluence of 1 mJ/cm$^2$ is shown in Fig. 3(a). The scaled peak shift of the fs data collapses into the synchrotron data of the very same sample after calibrating in consideration of their fluence difference, as shown in Fig. 3(b). We also notice that the maximum photoinduced strain at about $\tau$=12.5 ps is about 1.4 times larger than that at 100 ps. The dynamics can be interpreted as due to the propagation of coherent acoustic phonons. The delay $\tau$=12.5±1 ps corresponds to the time that the strain front travels once across the film with a velocity of $v=d/T$=3.44 km/s. The sound-speed limited strain dynamics within 100 ps implies an instantaneous stress upon photoexcitation.

To find the atomistic origin of the giant lattice response, we simulated the photostriction effect with the $\Delta$SCF method [18] based on the effect of photoexcited electron-hole pairs. The method fixes the occupations of certain states to take some electrons from the valence band to the conduction band and has been used to study the photostriction in ferroelectric and multiferroic structures [19, 20].

The density functional theory (DFT) code Abinit (v8.6) [21] was employed to calculate the lattice and electronic properties. Plane waves with the energy cutoff of 30 Hartree (816 eV) are used as the basis set. The GGA-PBEsol [22] functional and projector augmented wave (PAW) [23] potentials from the JTH dataset (v1.1) [24] are used. A Hubbard-U correction [25] with U(Co) =3.5 eV, which reproduces the *d-d* and *p-d* transition optical spectrum, is applied [3]. An $8\times8\times3$ non-shifted k-point grid is used to integrate the Brillouin Zone. The SCO structure has the Ima2 space group and G-type anti-ferromagnetic order. The structures are relaxed until the residual forces are below $1\times10^{-6}$ Ha/Bohr ($5.14\times10^{-5}$ eV/A). The occupations of a pair of bands in valence and conduction bands were fixed to 1-$n_e$ and $n_e$, respectively, to mimic the electron-hole pair with density of $n_e$. We consider the density in the dilute limit to keep the band structure almost unchanged so that it can be considered as a perturbation. Densities of charge transfer ranging from 0 to about $2\times10^{20}$ cm$^{-3}$ were considered. We did not choose any specific k-point for the electron transfer but fixed the occupation of two bands in the whole Brillouin zone. We then relaxed the structures with the band occupations fixed.

We selected a few bands to represent the (I) O 2*p*, (II) Co 3*d* (at CBM), (III) Co$^T$ 3*d*, and (IV) Co$^O$ 3*d*, with energies levels of about -1.5 eV, -0.2 eV, 1.0 eV, and 1.5 eV relative to the Fermi energy, respectively (Fig. 4(a)). The II→III transfer corresponds to the *d-d* transfer. Two kinds of *p-d* transfer: O 2*p* to 3*d* in Co$^O$ (I→III) and Co$^T$ (I→IV) are considered, respectively.

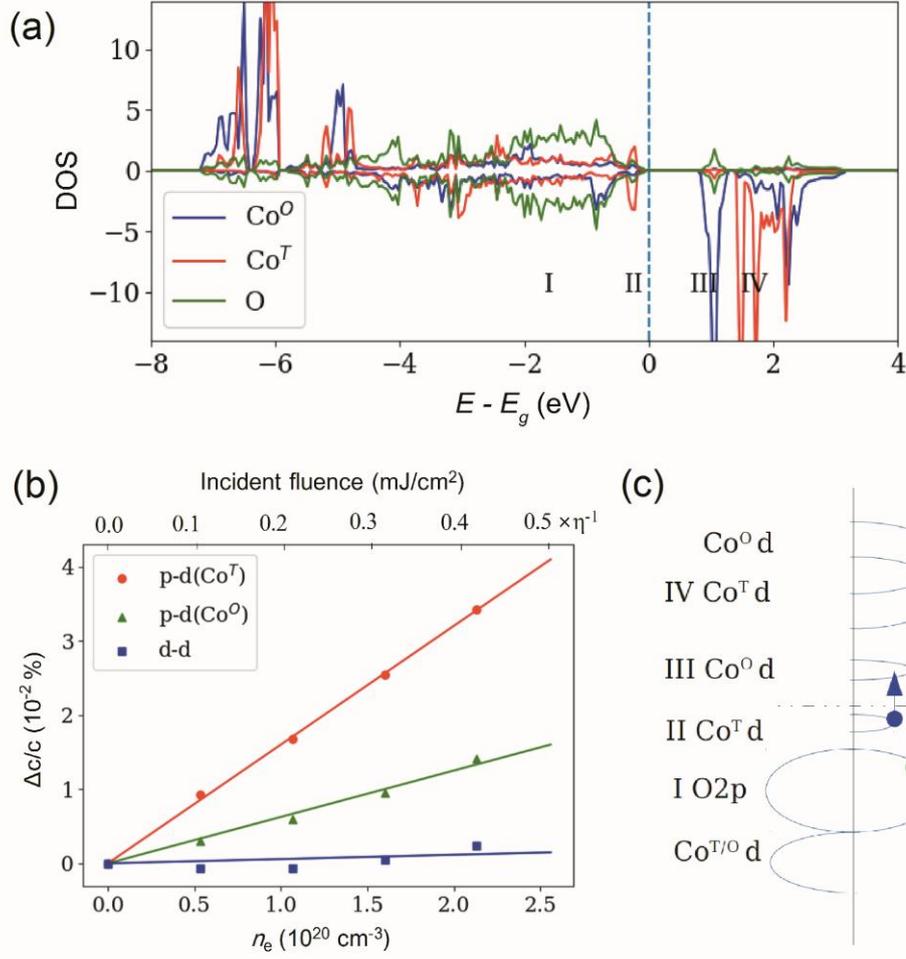

FIG. 4. (a) The density of states projected on Co 3d and O 2p orbitals. (b) The expansion of c-axis lattice versus the density of electron-hole pairs and the schematic transfers.

The result in Fig.4 (b) shows there is a distinct expansion of $c$ axis with $p$–$d$ charge transfer, whereas no significant change is found with $d$-$d$ charge transfer, which can explain the strong experimental photon energy dependence of photostriction shown in Fig. 2(a). Moreover, the linear trend of lattice change versus the concentration of excited electrons is also in good agreement with our fluence-dependent photoinduced strain data in Fig. 2(c). Quantitatively, assuming that the expansion is linear to the density of electron-hole pairs, it would take about $n_e=2\times10^{21}$ cm$^{-3}$ to achieve the huge expansion of 0.4%. Given the fluence that accounts for the 0.4% expansion in our UXRD measurements (absorbed fluence ~ 0.87 mJ/cm$^2$), the corresponding excited electron density can be easily estimated by $n_e^* = \eta \times F_{abs}/(E_p \times V)$ = $0.85\eta \times 10^{21}$ cm$^{-3}$, where $E_p$, $V$, $F_{abs}$ and $\eta$ refer to pump photon energy, pump volume, absorbed fluence and quantum efficiency, respectively. The estimated $n_e^*$ is slightly smaller than the predicted $n_e$, which seems to be reasonable given the uncertainty in determining the actual laser spot size and the assumption of the DFT calculation, e. g. density in dilute limit and the homogeneous charge transfer in Brillouin zone.

We first consider the thermal contribution and deformation potential part of the photostriction. The temperature jump is determined to be 90K [see Support Material] at most for the incident fluence of 1.55 mJ/cm$^2$, in case that all the photon energy transfer into heat. The corresponding thermal induced expansion is estimated as to be 0.18%, still far to explain the actual photoinduced strain of 0.404%. The distribution changes of the excited electron-holes might also contribute to the stress according to the deformation potential effect [26]. However, this can only lead to a negligible compression due to the negative sign of *dE$_g$/dP* (~ -0.04eV/GPa, [10]), contrary to our giant expansion.

Another possibility is the screening of internal electric field by mobile electrons. Though SCO is not ferroelectric, due to the layered $(Co^TO)^+$- (SrO) - $(Co^OO_2)^-$ - (SrO) structure, there is an electric field pointing from $(Co^T O)^+$ layers to $(Co^OO_2)^-$ layers (see Fig. S3). The $Sr^{2+}$ ions are attracted by the $(Co^OO_2)^-$ layer, reducing the distance of Sr-Sr with $Co^O$ (denoted as $l_O$, the other Sr-Sr distance denoted as $l_T$, labeled in Fig. S3). The Sr-Sr distances along the c-axis are ideal to detect the change of the field, since the screening of the field would reduce $l_O$ and increase $l_T$. In this scheme, the *p-d*($Co^O$) and *p-d*($Co^T$) charge transfer will surely exert opposite influence on the lattice response. However, we found that both of the two transfers indicate an expansion of c-axis (Fig. 4c). Therefore, it is unlikely that the screening of the electric field is the reason for the c-axis expansion.

In the light of the DFT simulation, it is more likely that the effect originates from the formation of antibonding state of the O *2p* and Co *3d* ($Co^T$) that lead to the expansion of the Co-O bonds. The states we referred to as O *2p* and Co *3d* correspond to the bonding and antibonding states rather than the atomic orbitals. The electrons occupying the antibonding state would reduce the energy gain by forming the Co-O bonds, leading to the bond expansion. This is consistent with the dependence on orbitals of electron-hole pairs. Indeed, we found that with electron transfer to $Co^O$ ($Co^T$), the $Co^O$-O ($Co^T$-O) bonds expand, as shown in Tab. 1. The increasing of Co-O-Co bond angle along c direction also contribute to the expansion of the c-axis with the *p-d* ($Co^T$) charge transfer.

Table I. Bond lengths and bond angles. The unit of bond lengths is Å; unit of bond angle is degree. All the bond lengths are along c-direction. The results were calculated with $2.2 \times 10^{20}$ cm$^{-3}$

|  | $l$ (Co$^T$-O) | $l$ (Co$^O$-O) | Co$^T$-O-Co$^O$ | $l^T$ (Sr-Sr) | $l^O$ (Sr-Sr) |
| --- | --- | --- | --- | --- | --- |
| Pristine | 1.8210 | 2.1684 | 154.186 | 4.3417 | 3.3978 |
| $p$-$d$ (Co$^O$) | 1.8207 | 2.1696 | 154.126 | 4.3436 | 3.3970 |
| $p$-$d$ (Co$^T$) | 1.8237 | 2.1654 | 154.382 | 4.3413 | 3.4001 |

We also calculated the structure of SCO with high, low, and intermediate spin state to see whether the spin crossover effect can lead to a c-axis expansion. A large c-axis elongation of about 0.4% can be found in the intermediate spin state due to the strong Jahn-Teller effect of the $d^4 \uparrow d^2 \downarrow$ Co$^O$ (see Table. S2 of the supplemental material). The energy cost to flip one spin is about 1.05eV, which is smaller than the photon energy, therefore the intermediate spin state could be reached through photon-excitation. Though the saturation value of c-axis expansion due to the spin-crossover is smaller than the experimental result, the high spin to intermediate spin crossover of Co ions could contribute to the huge lattice expansion.

In summary, we combine the multi-timescale UXRD experiments and DFT computation to study the photoinduced strain in an oxygen-ordered SCO film. Giant and strong photon energy dependent lattice distortion has been observed and is attributed to photoexcitation of bonding to antibonding state charge transfer, resulting in the bond extension in SCO unit cell. The photoinduced high spin to intermediate spin crossover is also not precluded. The giant lattice distortion observed in SCO might also be potentially generic for other oxides with oxygen vacancy because of their large tetragonality and the manipulation of the super-tetragonality via photoexcited carriers will also offer a promising route to tailoring the material properties and functionalities such as piezoelectricity or electrostriction.

This work was supported by the National Natural Science Foundation of China (Grant No. 11574365). The authors thank beamline 1W2B (Beijing Synchrotron Radiation Facility) and BL14B1 (Shanghai Synchrotron Radiation Facility) for providing the beam time and helps during experiments. The use of the Advanced Photon Source is supported by U.S. Department of Energy, Office of Science, Basic

Energy Sciences, under Contract No. DE-AC02-06CH11357. EB and XH acknowledge the ARC project AIMED and the F.R.S-FNRS for funding. EB and XH has relied on the CECI facilities funded by F.R.S-FNRS (Grant No. 2.5020.1) and Tier-1 super-computer of the Federation Wallonie-Bruxelles funded by the Walloon Region (Grant No. 1117545) for simulations.